\documentclass{article}
\usepackage{graphicx}
\usepackage{amsmath}

\begin{document}

%%*********************** TITLE **********************

\title{
\begin{flushright}
{\small USACH-FM-02/xx}\\[1.0cm]
\end{flushright}
{\bf Algebra of chiral currents on the physical surface}}

%%*********************** AUTHORS AND FILIATION **********************

\author{{ Alexis Am\'{e}zaga}\thanks{
E-mail: aamezaga@lauca.usach.cl}
{\ and Carlos Leiva}\thanks{
E-mail: caleiva@lauca.usach.cl}
\\
{\small {\it Departamento de F\'{\i}sica,
Universidad de Santiago de Chile,
Casilla 307, Santiago 2, Chile}}}
\date{}

\maketitle

\vskip-1.0cm

%********************** ABSTRACT ***********************

\begin{abstract}
\noindent Using a particular structure for the Lagrangian action
in a one-dimensional Thirring model and performing the Dirac's
procedure, we are able to obtain the algebra for chiral currents
which is entirely defined on the constraint surface in the
corresponding hamiltonian description of the theory.

\end{abstract}

%********************** SECTION 1 ***********************

\section{\protect\bigskip Introduction}

The Thirring model is a well known model in quantum field theory 
because it is exactly solvable in (1+1) dimensions \cite{Thirring58}. 
If this model is treated in a Hamiltonian way in terms of currents, with
interaction between them, it can be easily solved by performing a Bogoliubov 
transformation. This is, concerning to find the fermionic correlators using 
a bosonization procedure \cite{Stone94}.

An extensive investigation of its current algebra has been performed.
Dell' Antonio et. al. \cite{Antonio72}, using an exact expansion
in bilocal operators products, have solved the model entirely in terms
of currents. They defined the so-called Schwinger term as function of a
c-number and then determined their form requiring that a spinor
transformation law was fulfilled. With a current regularization, 
Takahashi and Ogura \cite{Ogura01} also obtained a very similar Schwinger 
term in their formulation of Thirring model.

A particulary interesting description of the current algebra of this
model is given by Gomes et. al. \cite{Gomes88}. Following a Dirac's
procedure, they are able to calculate the current algebra that
lives in the constraint surface. However, the inclusion in that surface
of the Hamiltonian which governs the dynamics and participates in the
algebra it is not assured in their procedure.

The purpose of this paper was, taking advantage of the special
Lagrangian action introduced by Floreanini and Jackiw for
self-dual fields \cite{Jackiw87} (used also by Gomes et. al.
\cite{Gomes88, Costa88, Girotti88}), we shall to analyze the
current algebra in a massless Thirring model where the proposed
Lagrangian action for the currents generates a Hamiltonian that is
defined also in the constraint surface. In this way, the current
algebra obtained by the Dirac's procedure \cite{Dirac64} entirely
lives in the same physical surface.

The plan of this paper is the following. In section 2, we review
the known current algebra in the Dirac formalism \cite{Gomes88},
and also obtain the corresponding Hamiltonian on the constraint
surface, which coincides with the usual form for the free theory.
In section 3, we repeat the procedure that was successful in the
free case, obtaining a complete interacting current algebra
defined on the physical surface. Finally in section 4, we examine
a formal extension of this procedure for the case of $N$
fermion currents with interactions, and in order to illustrate, we
solve the complete algebra for the $N=2$ case, which exhibits all
the interesting features claimed above.

%\bigskip

%********************** SECTION 2 ***********************

\section{Non-interacting Kac-Moody $U(1)$ algebra of chiral currents on the physical surface}

%\bigskip

%\qquad

Based on the treatment of fermion fields in terms of currents
explored by Dashen and Sharp \cite{Dashen68}, Sugawara
\cite{Sugawara68} and Sommerfield \cite{Sommerfield68}, let us
start with the following Lagrangian for the right ($J_{R}$) an
left ($J_{L}$) currents of the model

%\bigskip

\begin{equation}
L=\frac{1}{2}\int dxdy\left[ J_{R}(x)\varepsilon (x-y)\partial
_{+}J_{R}(y)-J_{L}(x)\varepsilon (x-y)\partial _{+}J_{L}(y)\right],
\label{2.1}
\end{equation}
\bigskip

\noindent where $J_{R,L}$ are the right and left currents and
$\varepsilon $ is the Heaviside function ($\varepsilon(0) \equiv 0$ 
is assumed). This Lagrangian is similar to the one used in
\cite{Jackiw87} and \cite{Gomes88}, but here the canonical
Hamiltonian is zero. Later a new one will be obtained on the
constraint surface, which satisfies the same motion equations.

%\bigskip

The constraints for this system are

\begin{eqnarray}
\chi _{1}(x) &=&\Pi _{R}(x)-\frac{1}{2}\int dy J_{R}(y)\varepsilon (y-x), \label{2.2}  \\
\chi _{2}(x) &=&\Pi _{L}(x)+\frac{1}{2}\int dy J_{L}(y)\varepsilon (y-x). \label{2.3}
\end{eqnarray}
\bigskip

These are second class constraints which are automatically conserved because the canonical Hamiltonian vanishes.

Following the Dirac's procedure we calculate the Dirac's brackets defined as

\begin{equation}
\left\{ A,B\right\} _{D}=\left\{ A,B\right\} -\int dzd\omega\left\{ A,\chi _{i}(z)\right\}
C_{i,j}^{-1}(z,\omega)\left\{ \chi _{j}(\omega),B\right\}, \nonumber
\end{equation}
\bigskip

\noindent where $C_{i,j}^{-1}$ is the constraint Poisson bracket matrix

\begin{equation}
C_{i,j}(z,\omega)= \left(
\begin{array}{cc}
1 & 0 \\
0 & -1
\end{array}
\right)
\epsilon(z-\omega), \label{2.3a}
\end{equation}
\bigskip

\noindent with

\begin{equation}
C_{i,j}^{-1}(z,\omega)= \left(
\begin{array}{cc}
-1 & 0 \\
0 & 1
\end{array}
\right)
\delta '(z-\omega), \label{2.3b}
\end{equation}
\bigskip

\noindent and $\chi _{i}$ denotes second class constraints; $ \left \{ \cdot ,\cdot \right\} $
sets for usual Poisson's brackets. After a simple calculation we obtain the algebra of currents
in this formalism

\begin{equation}
\left\{ J_{L}(x),J_{L}(y)\right\} _{D}= \delta '(x-y),    \label{2.4}
\end{equation}

\begin{equation}
\left\{ J_{R}(x),J_{R}(y)\right\} _{D}=-\delta '(x-y),  \label{2.5}
\end{equation}

\begin{equation}
\left\{ J_{L}(x),J_{R}(y)\right\} _{D}=0.  \label{2.6}
\end{equation}
\bigskip

This is the result given by Gomes et. al. in \cite{Gomes88}. Now
we want to obtain a similar Hamiltonian, but assuring by
construction \cite{Dirac64, Claudio91}, that it entirely lives on
the constraint surface, and that it governs the dynamics of
currents involved in the algebra (\ref{2.4})-(\ref{2.6})

To obtain the Hamiltonian we introduce the following generalized vector

\begin{equation}
Z^{\alpha}=\left\{ \Pi _{L}; \Pi _{R}; J_{L}; J_{R}\right\},
\label{2.7}
\end{equation}
\bigskip

\noindent on the constraint surface. As constraints themselves are
strong identities, we can parameterize the momenta in terms
of currents. Furthermore, in order to retrieve Hamiltonian
relations between our new variables, we must define a new
Hamiltonian on that surface according to

%\bigskip
\begin{equation}
K=H_{0}+F,  \label{2.8}
\end{equation}
\bigskip

\noindent where $H_{0}$ is the initial canonical Hamiltonian and
$F$ obeys the relationship

\begin{equation}
\frac{\delta F(x)}{\delta u^{i}}=\int dwdy\left( \frac{\partial Z^{\alpha
}(x)}{\partial u^{i}(w)}W_{\alpha \beta }(x,y)\frac{\partial Z^{\beta
}(y)}{\partial w}\right),  \label{2.9}
\end{equation}
\bigskip

\noindent where $W_{\alpha ,\beta }$ is the initial symplectic matrix, $u^{i}$ are $%
J_{L}$ or $J_{R}$, and $Z^{\alpha}$ is a component of $Z$ defined
in (\ref{2.7}). It looks like

\begin{equation}
W_{\alpha \beta}(x-y)=\left(
\begin{array}{cccc}
0 & 0 & \delta (x-y) & 0 \\
0 & 0 & 0 & \delta (x-y) \\
-\delta (x-y) & 0 & 0 & 0 \\
0 & -\delta (x-y) & 0 & 0
\end{array}
\right).  \label{2.9}
\end{equation}
\bigskip

The consistency of this procedure is ensured because $W$ is time independent.

In order to obtain $F$, we must observe its variations along $%
J_{L} $ and $\ J_{R}$

\begin{eqnarray}
\int dx\delta F(x) &=&\int dxdz(\left\{ F(x),J_{L}(z)\right\}
\delta
J_{L}(z)+\left\{ F(x),J_{R}(z)\right\} \delta J_{R}(z))  \nonumber \\
&=&\int dxdzdw\left( \frac{\partial F(x)}{\partial J_{L}(w)}\left\{
J_{L}(w),J_{L}(z)\right\} \delta J_{L}(z) \right.  \nonumber \\
&&\left. +\frac{\partial F(x)}{\partial J_{R}(w)}(z)\left\{ J_{R}(w),J_{R}(z)\right\} \delta J_{R}\right)  \nonumber \\
&=&-\int dxdz\partial _{z}\left( \frac{\partial F(x)}{\partial J_{R}(z)}%
\delta J_{R}(z)-\frac{\partial F(x)}{\partial J_{L}(z)}\delta J_{L}(z)\right) \nonumber \\
&=&\int dx\left( \frac{\partial F(x)}{\partial J_{L}(z)}\delta J_{L}(z)-%
\frac{\partial F(x)}{\partial J_{R}(z)}\delta J_{R}(z)\right). \label{2.10}
\end{eqnarray}
\bigskip

Note that from  (\ref{2.8}) and (\ref{2.9}) is easy to
calculate $\partial F/\partial J_{L}$ and $\partial F/\partial
J_{R}$, then

\begin{equation}
\int dx\delta F(x)=2\int dx[J_{L}(x)\delta J_{L}(x)-J_{R}(x)\delta J_{R}(x)].   \label{2.11}
\end{equation}
\bigskip

Therefore, due to the vanishing of the canonical Hamiltonian, the modified Hamiltonian is

%\bigskip

\begin{equation}
K=\int dx(J_{L}^{2}(x)+J_{R}^{2}(x)).  \label{2.12}
\end{equation}
\bigskip

Finally,

\begin{equation}
\left\{K, J_{L,R}(x) \right\}_{D}= \pm 2 J'_{L,R}(x).  \label{2.13}
\end{equation}
\bigskip

The advantage of this procedure is that the Hamiltonian that
participates in the Kac-Moody algebra is defined on the constraint
surface, i.e. on the physical surface, which guarantees that the
solution of motion equations is in complete consistency with the
algebra. On the other hand, this is very important because it assures
that in the quantization procedure of the theory, only gauge
degrees of freedom are counted.

%\bigskip

%********************** SECTION 3 ***********************

\section{Chiral current algebra on the physical surface for Thirring model}

%\bigskip

Now consider the Lagrangian

\begin{eqnarray}
L &=& \frac{1}{2}\int dxdy\left[ (J_{R}(x)-gJ_{L}(x))\varepsilon (x-y)\partial_{+}J_{R}(y)\right.  \nonumber \\
&&\left.-(J_{L}(x)+gJ_{R}(x))\varepsilon (x-y)\partial _{+}J_{L}(y) \right],  \label{3.1}
\end{eqnarray}
\bigskip

\noindent $g$ being the coupling constant. This one admit the same considerations
made in the previous section.

Now, our second-class conserved constraints are

%\bigskip

\begin{eqnarray}
\chi _{1} &=&\Pi _{R}(x)-\frac{1}{2}\int dy(J_{R}(y)-gJ_{L}(y))\varepsilon
(y-x),   \label{3.2} \\
\chi _{2} &=&\Pi _{L}(x)+\frac{1}{2}\int dy(J_{L}(y)+gJ_{R}(y))\varepsilon
(y-x).   \label{3.3}
\end{eqnarray}

\bigskip

These are again second class constraints and, as in the non-interacting case,
they are conserved (in this sense, it is a closed system).

Now we can follow the Dirac's procedure and calculate the Dirac's
brackets on the constraint surface. To do this, it is necessary to build $W$
(that is, the matrix that resumes the former bracket relations). We have

\begin{equation}
C(z)=\left(
\begin{array}{cc}
 1& -g \\
-g & -1
\end{array}
\right)\varepsilon (z).  \label{3.4}
\end{equation}
\bigskip

Note that this matrix is symmetric. In order to calculate Dirac's
brackets, we first need to calculate the inverse of $C$. We obtain

\begin{equation}
C^{-1}(z)=\frac{1}{(1+g^{2})}\left(
\begin{array}{cc}
1 & -g \\
-g & -1
\end{array}
\right)\partial _{z}\delta(z).  \label{3.5}
\end{equation}
\bigskip

The Dirac's brackets between currents on the constraint surface turn out to be

\begin{equation}
\left\{ J_{L}(x),J_{L}(y)\right\} _{D}=\frac{1}{(1+g^{2})} \delta '(x-y),  \label{3.6}
\end{equation}

\begin{equation}
\left\{ J_{R}(x),J_{R}(y)\right\} _{D}=\frac{-1}{(1+g^{2})} \delta '(x-y),  \label{3.7}
\end{equation}

\begin{equation}
\left\{ J_{L}(x),J_{R}(y)\right\} _{D}=\frac{-g}{(1+g^{2})} \delta '(x-y),  \label{3.8}
\end{equation}

\begin{equation}
\left\{ J_{R}(x),J_{L}(y)\right\} _{D}=\frac{-g}{(1+g^{2})} \delta '(x-y).  \label{3.9}
\end{equation}
\bigskip

Now, if we calculate the new Hamiltonian on the physical surface
by the same way as before, we arrive to

%bigskip

\begin{equation}
K=\int dx[J_{L}^{2}(x)+J_{R}^{2}(x)+2gJ_{L}(x)J_{R}(x)].  \label{3.10}
\end{equation}
\bigskip

So,

\begin{equation}
\left\{K, J_{L,R}(x) \right\}_{D}= \pm
\frac{2}{(1+g^{2})} \left[ J'_{L,R}(x)%
+ g J'_{R,L}(x) \right]. \label{2.13}
\end{equation}
\bigskip

We have obtained a modification of the Schwinger term as in
\cite{Antonio72, Ogura01}. It is remarkable that this result
involves the last section, which is reached directly vanishing
the coupling constant $g$.

%\bigskip

%%%********************** SECTION 4 ***********************

\section{An extension of the model}

The previous results can be extended straightforwardly to the case of $N$
currents with interactions among all them. Let us to consider the following Lagrangian

\begin{eqnarray}
L &=&\frac{1}{2}\int dxdy\sum_{i=1}^{N}\left[ J_{i}^{L}(x)-%
\sum_{j=1}^{N}g_{ij}J_{j}^{R}(x)\right] \varepsilon (x-y)\partial
_{+}J_{i}^{L}(y)  \nonumber \\
&&-\frac{1}{2}\int dxdy\sum_{i=1}^{N}\left[ J_{i}^{R}(x)+%
\sum_{j=1}^{N}g_{ij}J_{j}^{L}(x)\right] \varepsilon (x-y)\partial
_{+}J_{i}^{R}(y),  \label{4.1}
\end{eqnarray}
\bigskip

\noindent where $g_{ii}\equiv g$  and $g_{ij}\equiv h$ $%
\left( i\neq j\right) $. The currents are defined by

\begin{equation}
J_{i}^{L,R}(x)=\psi _{L,R}^{i^{\dagger }}(x)\psi _{L,R}^{i}(x).  \label{4.2}
\end{equation}
\bigskip

\noindent and the new constraint set is

\begin{equation}
\chi _{k}^{L}(z)=\Pi _{k}^{L}(z)-\frac{1}{2}\int dx\left[ J_{i}^{L}(x)-%
\sum_{j=1}^{N}g_{kj}J_{j}^{R}(x)\right] \varepsilon (x-z),  \label{4.3}
\end{equation}

\begin{equation}
\chi _{k}^{R}(z)=\Pi _{k}^{R}(z)+\frac{1}{2}\int dx\left[ J_{i}^{R}(x)+%
\sum_{j=1}^{N}g_{kj}J_{j}^{L}(x)\right] \varepsilon (x-z),  \label{4.4}
\end{equation}
\bigskip

\noindent where $J$ and $\Pi $\ are canonical fields, subject to

\begin{equation}
\left\{ J_{i}^{L,R}(x),\Pi _{k}^{L,R}(y)\right\} =\delta (x-y)\delta _{ik}.   \label{4.5}
\end{equation}

\bigskip

For this case, the whole Kac-Moody algebra in Dirac's formalism is

%\bigskip

\begin{multline}
\left\{ J_{k}^{\alpha }(x),J_{m}^{\beta }(y)\right\} _{D}=
\left\{J_{k}^{\alpha }(x),J_{m}^{\beta }(y)\right\}  \\
-\int dzd\omega \left\{ J_{k}^{\alpha}(x),\chi _{\mu }(z)\right\} C^{\mu \nu }(z-\omega)\left\{ \chi _{\nu
}(\omega),J_{m}^{\beta }(y)\right\},  \label{4.6}
\end{multline}
\bigskip

\noindent where $C_{\mu \nu }=\left\| \left\{ \chi _{\mu },\chi _{\nu }\right\} \right\|$
and $\int dzC_{\mu \tau }(x-z)C^{\tau \nu }(z-y)=\delta(x-y)\delta _{\mu }^{\nu }$.

\bigskip

The current Hamiltonian on the current surface of this model is

\begin{equation}
H=\int dx\left\{ \sum_{i=1}^{N}\left[ \left(J_{i}^{L}(x)\right)^{2}+\left(J_{i}^{R}(x)\right)^{2}%
\right] +\sum_{i,j=1}^{N}g_{ij}J_{i}^{L}(x)J_{j}^{R}(x)\right\}.    \label{4.7}
\end{equation}
\bigskip

Give a compact expression for the inverse of a ($N\times N$) matrix in the 
more general case is a very complicated task,and it is not very illustrative 
to our main objective. Thus, we will consider the $N=2$ case, which allows to illustrate all
desired features of this extension.

In this case we have

\begin{equation}
C_{\alpha \beta }\left( z\right) =\left(
\begin{array}{cccc}
1 & -g & 0 & -h \\
-g & -1 & -h & 0 \\
0 & -h & 1 & -g \\
-h & 0 & -g & -1
\end{array}
\right) \varepsilon \left( z\right),  \label{4.8}
\end{equation}
\bigskip

\noindent and, after some calculations, we obtain

\begin{eqnarray}
\left\{ J_{k}^{L}(x),J_{m}^{L}(y)\right\} _{D} &=&f
\delta '(x-y)\left[ \left( 1+g^{2}+h^{2}\right) \left( \delta _{k1}\delta
_{1m}+\delta _{k2}\delta _{2m}\right) \right.  \nonumber \\
&&\left. -2gh\left( \delta _{k1}\delta _{2m}+\delta _{k2}\delta _{1m}\right)
\right],        \label{4.9}
\end{eqnarray}

\begin{eqnarray}
\left\{ J_{k}^{R}(x),J_{m}^{R}(y)\right\} _{D} &=&-f
\delta '(x-y)\left[ \left( 1+g^{2}+h^{2}\right) \left( \delta _{k1}\delta
_{1m}+\delta _{k2}\delta _{2m}\right) \right.  \nonumber \\
&&\left. -2gh\left( \delta _{k1}\delta _{2m}+\delta _{k2}\delta _{1m}\right)
\right],        \label{4.10}
\end{eqnarray}

\begin{eqnarray}
\left\{ J_{k}^{L}(x),J_{m}^{R}(y)\right\} _{D} &=&-f
\delta '(x-y)\left[ \left( g+g^{3}-gh^{2}\right) \left( \delta
_{k1}\delta _{1m}+\delta _{k2}\delta _{2m}\right) \right.  \nonumber \\
&&\left. +\left( h+h^{3}-g^{2}h\right) \left( \delta _{k1}\delta
_{2m}+\delta _{k2}\delta _{1m}\right) \right],        \label{4.11}
\end{eqnarray}

\begin{eqnarray}
\left\{ J_{k}^{R}(x),J_{m}^{L}(y)\right\} _{D} &=&-f
\delta '(x-y)\left[ \left( g+g^{3}-gh^{2}\right) \left( \delta
_{k1}\delta _{1m}+\delta _{k2}\delta _{2m}\right) \right.  \nonumber \\
&&\left. +\left( h+h^{3}-g^{2}h\right) \left( \delta _{k1}\delta
_{2m}+\delta _{k2}\delta _{1m}\right) \right],        \label{4.12}
\end{eqnarray}
\bigskip

\noindent where

\begin{equation}
f=\frac{1}{\left[ 1+(g-h)^{2}\right] \left[ 1+(g+h)^{2}\right] }.     \label{4.13}
\end{equation}
\bigskip

The Hamiltonian on the constraint surface is

\begin{equation}
K=\int dx\left\{ \sum_{i=1}^{2}\left[ \left(J_{i}^{L}(x)\right)^{2}+\left(J_{i}^{R}(x)\right)^{2}%
\right] +\sum_{i,j=1}^{2}g_{ij}J_{i}^{L}(x)J_{j}^{R}(x)\right\},        \label{4.14}
\end{equation}
\bigskip

\noindent and

\begin{eqnarray}
\left\{K, J^{L,R}_{m}(x) \right\}_{D} &=& \pm 2f \left\{ (1+g^{2}+h^{2}) \left[ J'^{L,R}_{1}(x)
+ gJ'^{R,L}_{1}(x)  + hJ'^{R,L}_{2}(x) \right] \right. \nonumber \\
&&  -2gh \left[ J'^{L,R}_{2}(x) + gJ'^{R,L}_{2}(x)  + hJ'^{R,L}_{1}(x) \right]  \nonumber \\
&& \mp (g+g^{3}-gh^{2})h \left[ J'^{R,L}_{1}(x) + gJ'^{L,R}_{1}(x)  + hJ'^{L,R}_{2}(x) \right] \nonumber \\
&&\left. \mp g(h+h^{3}-g^{2}h) \left[ J'^{R,L}_{2}(x) + gJ'^{L,R}_{2}(x)  + hJ'^{L,R}_{1}(x) \right] \right\}
\delta _{1m} \nonumber \\
&& \pm 2f \left\{ (1+g^{2}+h^{2}) \left[ J'^{L,R}_{2}(x) + gJ'^{R,L}_{2}(x)  + hJ'^{R,L}_{1}(x) \right] \right. \nonumber \\
&&  -2gh \left[ J'^{L,R}_{1}(x) + gJ'^{R,L}_{1}(x)  + hJ'^{R,L}_{2}(x) \right]  \nonumber \\
&& \mp (g+g^{3}-gh^{2})h \left[ J'^{R,L}_{2}(x) + gJ'^{L,R}_{2}(x)  + hJ'^{L,R}_{1}(x) \right] \nonumber \\
&&\left. \mp g(h+h^{3}-g^{2}h) \left[ J'^{R,L}_{1}(x) + gJ'^{L,R}_{1}(x)  + hJ'^{L,R}_{2}(x) \right] \right\}
\delta _{2m} \nonumber
\end{eqnarray}

\bigskip

These are all the relationship for current algebra for the $N=2$ case. Also, it is direct
to obtain the previous results by by sequentially turning off the constants $h$ and $g$ 
and setting $m=1$ (sections 3 and 2, respectively). So, in this sense, the consistency of 
this extension is fulfilled.

%
%
%%********************** CONCLUSIONS ***********************

\section{Conclusions}

In this contribution, a Lagrangian formulation was proposed for
the Thirring chiral current model. It allows to give a treatment
for the current algebra entirely on the physical surface, and this
algebra has the usual features reported by other authors. We also
give a first generalization of our results for $N$ currents with
interaction between them. The remarkable advantage of this
procedure is that the Hamiltonian as well as the others elements
that participate in the Kac-Moody algebra are defined on the
constraint surface. The fact that the Hamiltonian lives on the
constraint surface ensures that the equation of motion of the
theory has a full consistency with the Kac-Moody algebra that currents
satisfy. We consider this result very important because the corresponding 
quantized theory ("á la Dirac", for instance) is a gauge theory. 
The complete theory that lives on the physical surface ensures 
that the gauge symmetry will be warranted.

%%%********************** Acknowledges ***********************

\section*{Acknowledgments}
We would like to thank to J. Saavedra and S. Lepe for his useful observations
and comments. A. A. thanks that this work has been partially supported by the
MECESUP project No. 9903 from University of Santiago.

%*********************** REFERENCES ********************************************

%*******************************************************************

\end{document}